\documentstyle[aps,prb,multicol,epsf,picinpar]{revtex}

\def\e2 {\epsilon-\epsilon_k}
\def\be {\begin{equation}}
\def\ee {\end{equation}}
\def\bea {\begin{eqnarray}}
\def\eea {\end{eqnarray}}

\begin{document}
\draft
\title{Comment on cond-mat/0007299 by Schofield and Sandeman : inconsistency
with experiments }
\author{George Kastrinakis}
\address{Institute for Electronic Structure and Laser (IESL), Foundation for
Research and Technology - Hellas (FORTH), \\
P.O. Box 1527, Iraklio, Crete 71110, Greece}

\date{December 28, 2001}
\maketitle
\begin{abstract}

The inconsistency of cond-mat/0007299 and Phys. Rev. B 63 094510 (2001)
with ARPES and resistivity data is pointed-out. 

\end{abstract}

\vspace{0.7cm}

Schofield and Sandeman (SS) \cite{ss} presented a 'cold-spot'
model for the normal state 1-particle scattering rate and transport in the 
cuprates. They assume that the in-plane momentum $k$ dependent rate 
is given by their eq. (1)
\be
\Gamma = \Gamma_f \cos^2 2\theta + \Gamma_s \sin^2 2\theta \;,
\ee
where $\theta$ is the angle between $k$ and the a-axis crystal direction.

The resistivity in the SS model is given by
\be
\rho = a \sqrt{\Gamma_f \Gamma_s} \;\;, a=\text{const.}
\ee

SS do not offer a temperature $T$ dependence of 
$\Gamma_f$ and $\Gamma_s$ of their own. Instead, they resort
to mentioning as a possibility the Ioffe and Millis \cite{im} version of these
quantities, namely 
\be
\Gamma_f = \text{const.} \;\;, \;\Gamma_s \propto T^2.
\ee

The recent experimental facts are twofold. First, the ARPES expts.
\cite{valla}, which show a scattering rate quadratic in $x$, 
$x$=max\{$T$,energy\}, for $x\rightarrow 0$, but linear in $x$ if $x$ exceeds
a threshold. Second, the resistivity measurements at very low $T$ in
the normal state of BiSrLaCuO \cite{ono} (by use of strong pulsed
magnetic fields to supress the onset of superconductivity). 
The latter are consistent 
with a resistivity linear in $T$ down to low $T$, which saturates
for $T \rightarrow 0$.
In all, the ARPES 
and resistivity measurements give a consistent picture (although
they were carried out in different materials).

The SS results given by eqs. (1) and (2) {\em cannot}
reproduce in a consistent manner the ARPES and resistivity expts.
E.g., plugging eq. (3) into (1) yields a usual Fermi liquid rate
and a $T$-linear resistivity. On the contrary, the predictions 
of the microscopic model of refs. \cite{gk,gk2} are 
consistent with the aforementioned expts.

Interestingly, the experimental papers cited had 
appeared before the SS paper was submitted.

\vspace{.5cm}

Useful correspondence with Y. Ando is acknowledged.

\end{document}